\documentstyle[aps,prl,psfig,manuscript]{revtex}

\begin{document}
\draft

\title{Excitation Function of Energy Density and Partonic Degrees 
	of Freedom in Relativistic Heavy Ion Collisions
 } 

\author{H.~Weber, C.~Ernst, 
	M.~Bleicher\footnote{Fellow of the Josef Buchmann Foundation}, 
	L.~Bravina\footnote{Fellow of the Alexander v. Humboldt Foundation},
	H.~St\"ocker and W.~Greiner}
\address{
  	Institut f\"ur Theoretische Physik \\
	Johann Wolfgang Goethe Universit\"at\\
	Robert Mayer Str. 8-10\\
	D-60054 Frankfurt am Main, Germany}

\author{C. Spieles\footnote{Feodor Lynen Fellow of the Alexander v. Humboldt
	Foundation}}
\address{
	Lawrence Berkeley Laboratory\\
	1 Cyclotron Road\\
	Berkeley, CA 94720, USA
	}

\author{S.A. Bass$^\ddagger$}
\address{
	Department of Physics\\
	Duke University\\
	Durham, NC, 27708-0305, USA
	}

\maketitle

\begin{abstract}
We estimate the energy density $\epsilon$ pile-up at mid-rapidity 
in central Pb+Pb collisions from 2 -- 200 GeV/nucleon.
$\epsilon$ is decomposed into hadronic and partonic contributions.
A detailed analysis of the collision dynamics in the framework of a
microscopic transport model shows the importance of partonic degrees
of freedom and rescattering of leading (di)quarks in the early 
phase of the reaction for $E_{lab} \ge 30$~GeV/nucleon. 
In Pb+Pb collisions at 160 GeV/nucleon the energy density reaches up to
4 GeV/fm$^3$, 95\% of which are contained in partonic degrees of freedom.
\end{abstract}

\newpage

The determination of energy densities in ultra-relativistic heavy-ion
collisions is crucial for any discussion involving a possible deconfinement
phase transition to a QGP 
\cite{collins75a,stoecker80a,shuryak80a,mclerran86a,stoecker86a,clare86a,kajantie87a}.
Energy densities are not directly measurable -- therefore
a comparison between data and models must be used to pin down the
time evolution of the reaction. 
Then,  estimates for the energy density
during the hot and dense early reaction stage 
\cite{vanhove82a,bjorken83a,mclerran86a,stoecker86a,clare86a}
can be obtained from the most reasonably looking models.

It has been questioned whether hadronic transport models are still valid
at CERN/SPS energies: the energy density estimates obtained in these 
frameworks are claimed to be well above the critical energy density 
estimates for a deconfinement phase transition given by 
Lattice Gauge Theory \cite{blum95a,boyd95a,laermann96a}.
Hadronic transport models, however, contain implicit partonic
degrees of freedom \cite{sorge90a} -- particle production at high energies is 
e.g.\ modeled via the production and fragmentation of strings
\cite{andersson87a,andersson87b,sjoestrand94a}. This issue is
studied in the present paper.

In the UrQMD model used below, the leading
hadrons of the fragmenting string contain the valence-quarks of the
original excited hadron. These leading hadrons are allowed -- in the model -- 
to interact even during their formation time, with a reduced cross section, 
thus accounting for the original valence quarks contained in that
hadron. Those leading hadrons represent a simplified
picture of the leading (di)quarks of the fragmenting string. 
Newly to-be-produced hadrons which do not contain string valence quarks
do in the present model not interact during their formation time -- however,
they contribute to the energy density of the system.
A proper treatment of the partonic degrees of freedom during the
formation time ought to include soft and hard parton scattering
\cite{geiger92a} and the explicit time-dependence of the color interaction
between the expanding quantum wave-packets \cite{gerland98a}:
However, such an improved treatment of the internal hadron dynamics has
not been implemented for light quarks into the present model.
Therefore, in the following analysis all contributions stemming from
hadrons within their formation time are termed ``partonic''. 
All contributions stemming from fully formed hadrons are termed
``hadronic''. The main focus of this paper is on the partitioning
and  the time evolution
of the energy density and the collision
dynamics of the early, intermediate, and late reaction stage at 
energies $E_{lab} = 10 - 200$~GeV/nucleon.

For our analysis we employ the UrQMD model \cite{bass98a}, which is
based on analogous principles as 
(Relativistic) Quantum Molecular Dynamics
\cite{peilert88a,hartnack89b,aichelin91a,sorge89a,sorge95b,lehmann95a}.
Hadrons are represented by Gaussians   
in phase space and are propagated according
to Hamilton's equation of motion.
The collision term of the UrQMD model treats 55 different
isospin (T) degenerate baryon (B) species
(including nucleon-, delta- and hyperon- resonances with masses up to 2 GeV)
and 32 different T-degenerate meson (M) species,
including (strange) resonances as well as their
corresponding anti-particles, 
i.e.\ full baryon-antibaryon symmetry is included.
Isospin is treated explicitly.
For hadronic excitations with masses $m>2$ GeV (B) and $>1.5$ GeV (M)
a string model is used. Particles produced in the string fragmentation
are assigned a formation time. 
This time $\tau_f$ physically consists of a quantal time $\tau_Q$, 
i.e.\ before the partons are produced, $\tau_Q \sim 1/m$, and a quantum
diffusion time, $\tau_D$, during which the partons evolve in the
medium to build up their internal asymptotic wave-functions to form
the hadron. $\tau_Q$ and $\tau_D$ differ for different parton and hadron
species. For our present purpose, we -- for the sake of simplicity --
just collect all partons, formed and unformed, as one species.
For a detailed overview of the elementary cross sections and string excitation
scheme included in the UrQMD model, see ref. \cite{bass98a}.

\begin{figure}

\centerline{\psfig{figure=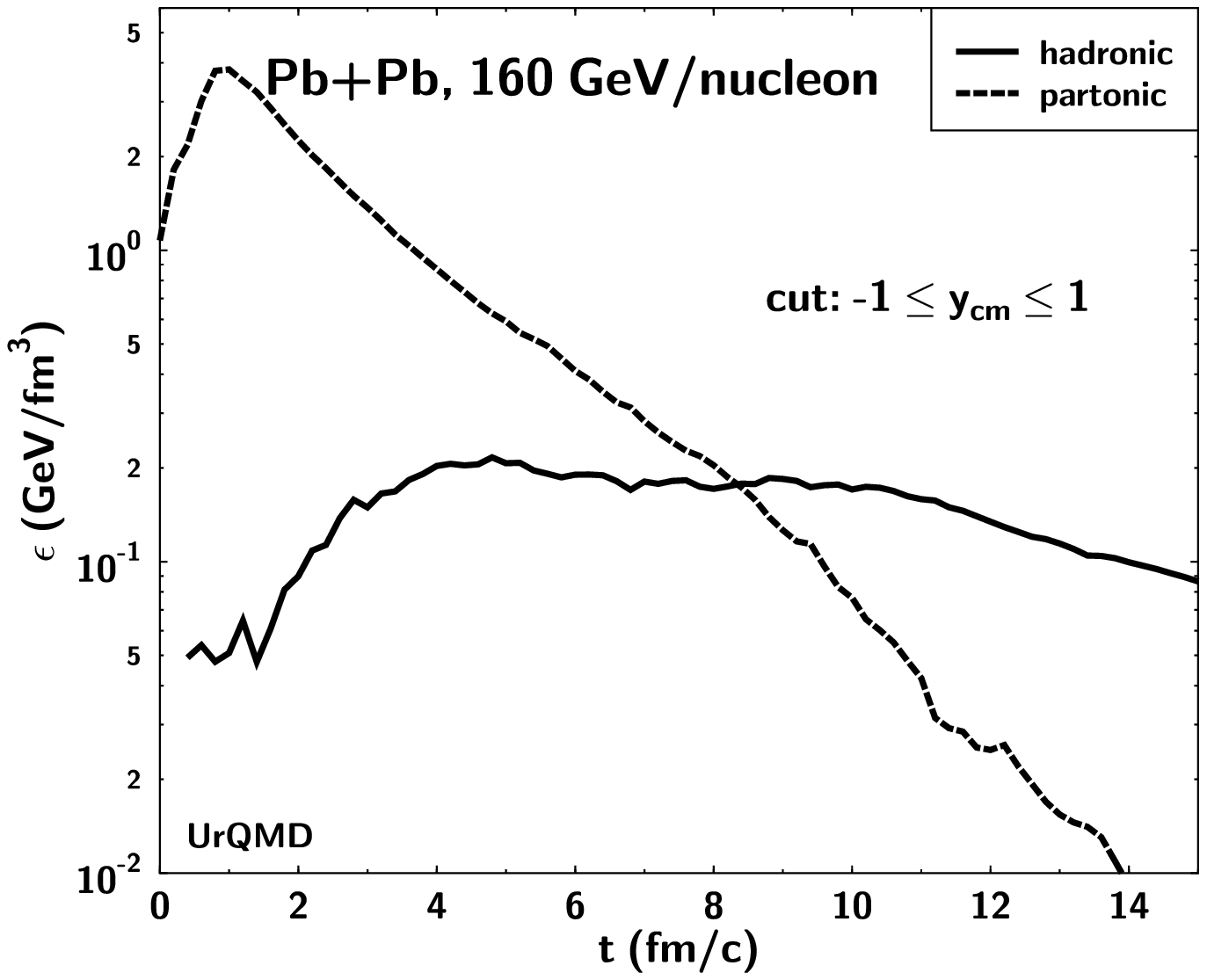,width=4in}}
\caption{\label{edens} Time evolution of the energy density $\epsilon$
	in central Pb+Pb reactions at 160 GeV/nucleon. $\epsilon$ has
	been decomposed into ``partonic'' and ``hadronic'' contributions
	and only particles around mid-rapidity have been taken into account.
	The early and intermediate reaction stages are dominated by
	the ``partonic'' contribution.}
\end{figure}

The partitioning of the distinct constituents can be inspected in 
figure \ref{edens} which shows the time-evolution of the energy 
density for central
Pb+Pb collisions at 160 GeV/nucleon. The nuclei are initialized such that
they touch a $t=0$ fm/c.
The energy density is partitioned into 
the above defined ``hadronic'' contribution, 
from fully formed hadrons, and the ``partonic'' contribution, 
from partons, constituent quarks and diquarks 
within the hadron formation time. 
Nearly all incident baryons are rapidly excited
into strings.  Subsequently,  ``partonic'' energy density builds up,
reaching values
of 4 GeV/fm$^3$ around midrapidity, $\Delta y = 1$ (at $t\approx 1$ fm/c). 
In the course of the reaction
hadrons are formed which increases in the ``hadronic''
energy density, accompanied by a nearly exponential decrease 
in the ``partonic'' energy density.

These energy densities are calculated as follows: In the UrQMD model hadrons
are represented by Gaussian wave packets. The width of the Gaussians
$\sigma=1.04$ fm and their normalization are chosen such that a calculation of 
the baryon density in the initial nuclei  yields ground state nuclear
matter density.
The (energy-) densities in the central reaction zone 
are obtained by summing analytically over all 
Gaussian  hadrons around mid-rapidity
($y_{c.m.} \pm 1$) at the locations of 
these hadrons and then averaging over these energy densities.
This summation over Gaussians  yields a smooth estimate
for baryon- and energy-densities, as compared to counting hadrons in a 
test volume. The rapidity cut insures that only those particles are taken
into account which have interacted. 
Thus, the  free streaming ``spectator'' matter is discarded.

The absolute value of the energy density, however, may depend on the
rapidity cut: Without rapidity cut the energy densities during the
early reaction stage ($t \approx 1$ fm/c) can be as high as 20 GeV/fm$^3$.
Even higher values in $\epsilon$ can
be obtained by choosing the geometric center of the collision
for the sum over the Gaussians instead of averaging over the
energy densities at the locations of the hadrons. The energy
density at a single point may not be physically meaningful
and therefore the latter method is favorable. 

Calculating the energy density by summing over particles 
 in the central reaction cell
may yield values for the total energy density of up
to 30 GeV/fm$^3$ (for a cell of $2\times 2 \times 1$ fm$^3$).
However, the values depend strongly on the volume of the cell.

\begin{figure}
\centerline{\psfig{figure=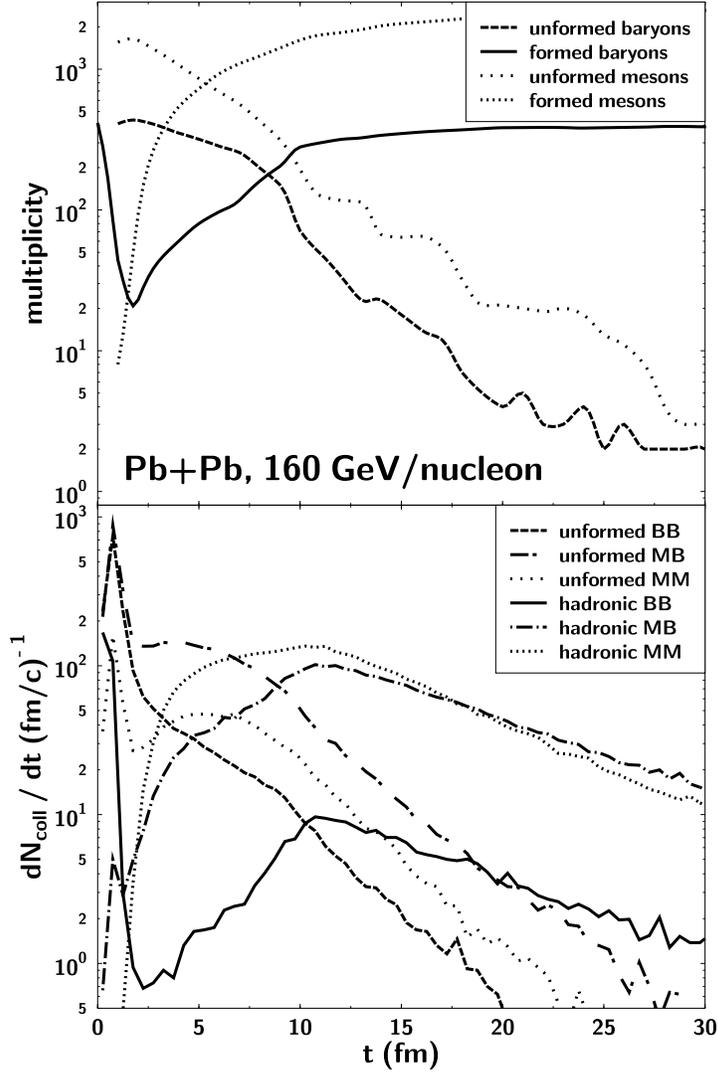,width=4in}}
\caption{\label{npart}
	Top: Time evolution of the multiplicity of hadrons and partonic
	constituents, divided into baryonic and mesonic contributions.
	Bottom: Collision rates for baryon-baryon (BB) and meson-meson (MM)
	collisions. The rates have been decomposed into interactions
	involving formed hadrons and those involving partonic constituents.}
\end{figure}

The time evolution of partonic constituents and  hadrons 
is shown in the upper frame of 
figure \ref{npart}. The first 5 fm/c of the 
reaction are dominated by the partonic constituents. 
The long-dashed and the dotted curves show the 
number of baryons and mesons contained in those constituents. 
In the case of leading-particles these can be interpreted as  
constituent (di)quarks or, for freshly born partons with small cross sections,
as excitation modes of the color field.

The lower frame of figure~\ref{npart}
shows the time evolution of the number of 
baryon-baryon (BB), meson-baryon (MB) 
and meson-meson (MM) collisions, both for ``hadronic''
and ``partonic'' interactions. ``Partonic'' interactions denote
interactions of {\em leading} (di)quarks either among themselves or with fully
formed hadrons. The early reaction stages, especially the
MM and MB cases, are clearly dominated by those ``partonic'' interactions. 
First after $\approx 10$ fm/c do collisions among fully formed hadrons
dominate the collision dynamics.
This number increases further if the scattering of the newly formed
partons is included. Thus 
``partonic'' degrees of freedom significantly contribute both,
to the energy density, as well as to the collision dynamics in the
first 5 fm/c. 

It should be noted that the "partonic"  collision
rates  strongly depend on the treatment of the partonic cross section
during formation time: In this analysis all interactions during
formation time have been considered purely "partonic". Other
scenarios, however, include a "hadronic" contribution 
to the cross section which increases continuously
during $\tau_D$ and reaches its full hadronic value at the end
of $\tau_D$ \cite{gerland98a}.

\begin{figure}
\centerline{\psfig{figure=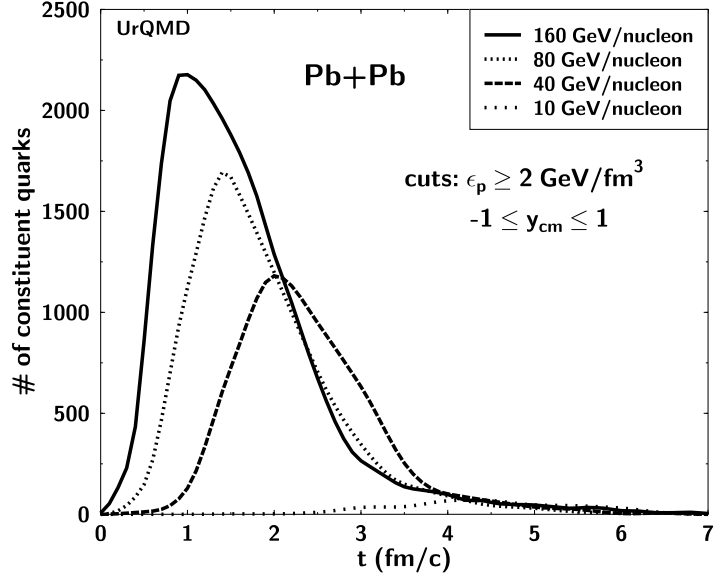,width=4in}}
\caption{\label{partmult}
	Multiplicity  of ``constituent quarks'' 
	propagating through a ``partonic'' energy density of 
	$\epsilon_Q \ge 2$ GeV/fm$^3$.
	The number of ``constituent quarks'' is obtained by summing over
	all partonic constituents, weighting the developing baryons
	by a factor of 3 and mesons-to-be by a factor of 2, respectively.
	At 160 GeV $>1000$ ``constituent quarks''
	are present over a time scale of $\Delta t \approx 2$ fm/c.}	
\end{figure}

The multiplicities of partons/hadrons actually contained in the region 
of high energy density are shown in figure~\ref{partmult}  
for different incident beam energies: 
We define this region to have a ``partonic'' energy density of 
$\epsilon_Q \ge 2$ GeV/fm$^3$ and
plot the  multiplicity 
of ``constituent quarks'' 
propagating through this energy density. 
The number of ``constituent quarks'' is obtained by summing over
all selected partonic constituents, weighting the developing baryons
by a factor of 3 and mesons-to-be by a factor of 2, respectively.
This allows to estimate the volume
of high ``partonic'' energy density: 
at 160 GeV, more than  1000 ``valence quarks''
are present over a time scale of $\Delta t \approx 2$ fm/c 
at $\epsilon_Q \ge 2$ GeV/fm$^3$.
Already at  beam energies around 40 GeV/nucleon, a sizable ``partonic''
phase exists; at 10.6 GeV/nucleon, however, the number of
``constituent quarks'' propagating through high ``partonic'' energy
density $\epsilon_Q \ge 2$ GeV/fm$^3$ is negligible.

However, during the early reaction stages matter in the central reaction
volume is neither fully hadronic, nor thermally and chemically equilibrated. 
A detailed analysis of velocity distributions and chemical composition
in a central cell for Au+Au reactions at 
10.6 GeV/nucleon \cite{bravina98a} shows that
first at times $t \approx 10$ fm/c hadronic matter in the central cell
can be viewed as nearly chemically equilibrated. This matter however never
exceeds energy densities of $\sim 1$ GeV/fm$^{-3}$, i.e.\ a density
above which the notion of separated hadrons loses its meaning.

\begin{figure}
\centerline{\psfig{figure=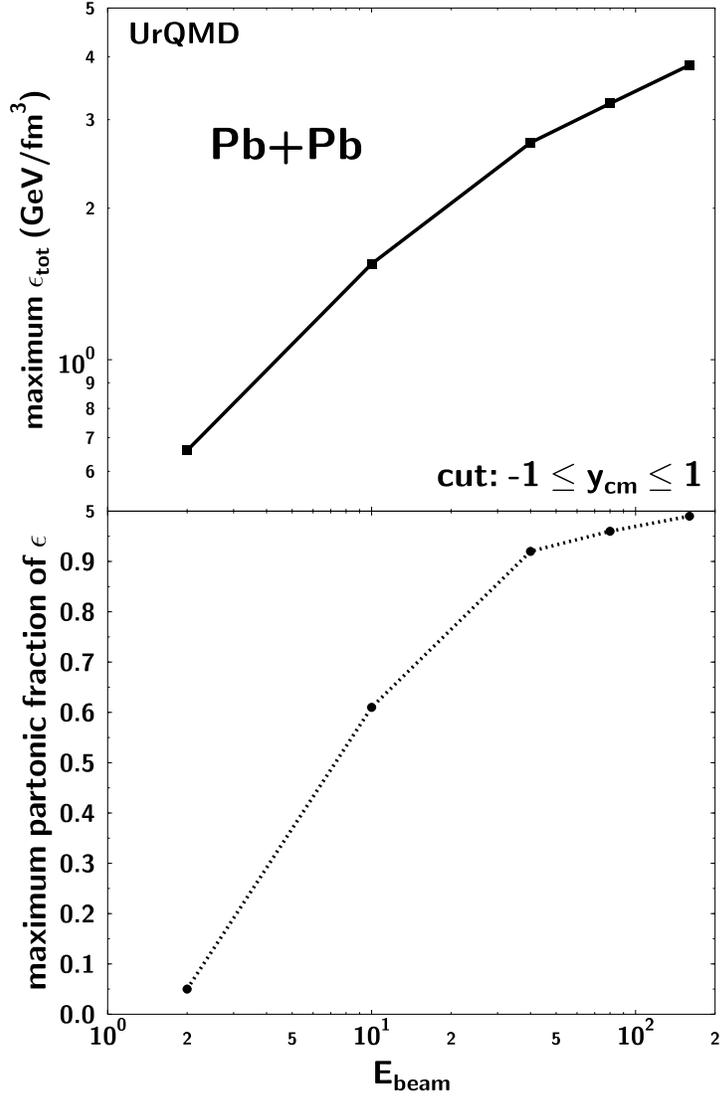,width=4in}}
\caption{\label{exfun}
	Top: excitation function of the maximum total energy density
	mid-rapidity hadrons experience. 
	Bottom: excitation function of the maximum ``partonic'' fraction
	of energy density. Already at a beam energy of 40 GeV/nucleon
	more than 90\% of the energy density is contained in partonic
	degrees of freedom at one time during the collision.
	}
\end{figure}

Do ``partonic''
degrees of freedom play any role at 10 GeV/nucleon, i.e.\ at the AGS?
The upper frame of figure~\ref{exfun} 
shows the maximum total energy density obtained in
central collisions of heavy nuclei as a function of incident beam energy,
starting from 2 GeV/nucleon and going  up to 200 GeV/nucleon. 
The energy density is obtained by the same method as used figure~\ref{edens}.
However, here ``partonic'' and ``hadronic'' contributions have been summed.
$\epsilon$ increases monotonously with the beam energy, 
reaching 
values as high as 4 GeV/fm$^3$ for SPS energies, which would seem
unreasonably high, if a purely hadronic 
scenario were used. 

The lower frame of figure~\ref{exfun} shows the maximum 
fraction of the energy density which is
contained in ``partonic'' degrees of freedom. Even at AGS
energies, already more than half of the energy density is due to
such ``partonic'' degrees of freedom, even though these do not yet dominate
the ``hadronic'' contributions.
At 40 GeV/nucleon, the maximum of the fraction of ``partonic'' energy
density is already $>90$\% of the total $\epsilon$.

The monotonous increase of the energy density 
excitation function does not imply that the excitation function of the
space-time volume of high {\em baryon density} shows the same behavior.
At AGS energies, $E_{lab} \sim 10$ GeV/nucleon, 
baryons still dominate the composition of the hadronic matter,
whereas at CERN/SPS energies, 200 GeV/nucleon, 
mesons constitute the largest fraction
of the hadronic matter. The maximum space-time volume of dense {\em baryonic}
matter can be reached at beam energies around 40 GeV/nucleon. A
detailed analysis of that regime, also with respect to experimental
signatures,  is presently underway \cite{weber98b}.

The importance of ``partonic'' degrees of freedom
for the collision dynamics in ultra-relativistic heavy ion collisions at
CERN/SPS energies does not imply that an equilibrated
Quark-Gluon-Plasma has been formed. In the UrQMD approach the ``partonic''
phase has been modeled as an incoherent superposition of non-interacting
partonic constituents. 
Furthermore, these ``partons'' retain their original correlation into
hadrons -- deconfinement is not implemented into the present UrQMD approach.
The leading (di)quark interactions (among each
other and with fully formed hadrons) constitute 
an interacting ``mixed phase'' (for the constituent parton dynamics in 
this model, see, however \cite{gerland98a,spieles97c}).
In contrast, parton cascades \cite{geiger92a,geiger97a} 
allow for interactions among the partons only, while 
hadronic final state interactions are to a large extent neglected.
Introducing the parton color dynamics, e.g. via an 
increasing cross section for partonic constituents
according to the QCD factorization theorem is a step towards such a
more complete scenario \cite{gerland98a}, including both parton- and
hadron rescattering, thus allowing ultimately also for a detailed
study of the equilibration of the constituents.

The question to be studied in such a model would be  
whether the partonic phase in relativistic heavy ion collisions
rather resembles more a  non-interacting free streaming parton gas
or a strongly interacting, nearly equilibrated 
``soup'' of partons, which so far has been the
prevalent simplified picture of a QGP.



S.A.B. acknowledges helpful discussions with Berndt M\"uller.
This work has been supported by GSI, BMBF, Graduiertenkolleg ``Theoretische
und experimentelle Schwerionenphysik'' DFG and DOE grant DE-FG02-96ER40945.

\end{document}